\documentstyle{article}
\begin{document}
\begin{center}
{\Large \bf Resonant peak splitting for ballistic conductance in
magnetic superlattices}
\end{center}

\vspace{.3cm}
\noindent
\begin{center}
{\large Z. Y. Zeng$^{1\ddagger}$, L. D. Zhang$^1$, X. H. Yan$^2$,
and J. Q. You$^2$}
\end{center}

\vspace{.001cm}
\noindent
\begin{flushleft}
{\it $^1 $Institute of Solid State Physics, Chinese Academy of
Sciences, \rm P.O. Box 1129, Hefei,\\ 230031, P. R. China \\
\it $^2$Department of
Physics, Xiangtan University, Xiangtan, Hunan 411105, P. R. China\\}
\end{flushleft}

\vspace{.5cm}
\noindent
\begin{center}
{\bf Abstract}
\end{center}

We investigate theoretically the resonant splitting of ballistic
conductance peaks in magnetic superlattices. It is found that, for
magnetic superlattices with periodically arranged $n$ identical
magnetic-barriers, there exists a general $(n-1)$-fold resonant
peak  splitting rule for ballistic  conductance, which is the
analogy of the $(n-1)$-fold resonant splitting for transmission in
$n$-barrier electric superlattices (R. Tsu and L. Esaki, Appl.
Phys. Lett. {\bf 22}, 562 (1973)).

\vspace{1.cm}
\noindent
{\bf PACS} numbers: 73.40.{\bf Gk}, 73.40.{\bf Kp}, 03.65.{\bf Ge}

\newpage

\noindent

Electron motion in a two-dimensional electron gas($2$DEG) subjected to a
magnetic
field has attracted long-drawn interest, since it provides a
variety of interesting and significant information characterizing the
behavior of
electrons in $2$DEG systems. There are a number of papers$^1$
devoted to the study on quantum transport of a $2$DEG in a unidirectional
weak sinusoidal
magnetic-field modulation with a uniform magnetic-field background,
where commensurability
effects come into play. This system was recently realized experimentally$^2$,
and
the long-predicted magnetoresistance oscillations$^1$ resulting
from semiclassical
commensurability between the classical cyclotron diameter and the period of
the magnetic modulation, were observed. Recently, a $2$DEG was investigated
under the influence of a magnetic step, magnetic well and magnetic
barrier$^3$.
Electron tunneling in more complicated and more realistic magnetic structures
was found to possess wave-vector filtering properties$^{4}$. The studies$^5$
showed that the energy spectrum of magnetic superlattice(MS)
consists of magnetic minibands.

    $(n-1)$-fold transmission splitting for $n$-electric-barrier
    tunneling was first noticed and generalized by Tsu and Esaki in
    their  pioneering paper$^6$, and was proved analytically by Lui and
    Stamp$^7$ in the electric superlattice(ES)
    with periodically arranged $n$ identical
    rectangular barriers. Very recently, Guo et al.$^{8}$
    investigated theoretically the
    transmission splitting effects in two kinds of
    magnetic supperlattices(MS) and found no explicit and general
    resonant peak splitting for transmission in electron tunneling
    in MS.

    We noticed that there is a single conductance peak for electron
    tunneling through the two-barrier magnetic structure, and two resonant
    spikes in the triple-barrier structure$^{4}$.   We also observed
    four resonant peaks in the ballistic conductance at low Fermi energies
    and  found that four resonant shoulders can be
    resolved
    for a $2$DEG modulated by a sinusoidal magnetic field of five
    periods and  a $5$-magnetic-step-barrier structure as long as
    the magnetic
    strength is strong enough$^{9}$. This urges us to explore
    whether there is a general resonant peak splitting rule
    for ballistic conductance in magnetic superlattices(MS).
    Since ballistic conductance
    can be derived as the electron flow averaged over half the
    Fermi surface$^{10}$, the main features of resonant tunneling
    through magnetic barriers is still preserved for ballistic
    conductance$^4$. It hints that there exists some kind of
    resonant peak splitting rule for ballsitic conductance
    in magnetic superlattices(MS), which is only dependent on the
    number of barriers in the magnetic field profile.
    In the following, we calculated ballistic conductances
    with the help of transfer matrix method in
     four kinds of magnetic superlattices$^{5}$:
     (a) Kronig-Penney magnetic superlattice
    (KPMS),
   which is the analogy of the well-known electrostatic Kronig-Penny model.
   Its magnetic field profile
   is modeled by the expression  $B(x)/B_0=
  \gamma\sum_{n=-\infty}^{n=+\infty}(-1)^{n+1}\delta(x-nl/2)$,
  and the vector potential can be taken as
  $A(x)/A_0=\gamma\frac12sgn[\cos(2 \pi x /l)]$(see Fig. $1$ (a)).
    (b) Step magnetic superlattice (Step MS):
    $B(x)/B_0=\gamma\sum_{n=-\infty}^{n=+\infty}
    (-1)^n\theta(x-nl/2)\theta[(n+1)l/2-x]$ and
   $ A(x)/A_0=\gamma\sum_{n=-\infty}^{n=+\infty}(-1)^n
    [x-(2n+1)l/4]\theta(x-nl/2)\theta[(n+1)l/2-x]$ (see Fig. $1$(b)).
    (c)Sinusoidal magnetic superlattice (Sinusoidal MS):
     $B(x)/B_0=\gamma\sin(2\pi x / l)$ and
   $A(x)/A_0=-\gamma\frac{l}{2\pi}\cos(2\pi x / l)$(see Fig. $1$(c)).
    (d) Sawtooth magnetic superlattice (Sawtooth MS):
    $B(x)/B_0=-\gamma\frac2l\sum_{n=-\infty}^{n=+\infty}
    (x-nl)\theta[x-(n+1/2)l]\theta(x-nl)$ and
    $A(x)/A_0=\gamma\frac1l\sum_{n=-\infty}^{n=+\infty}
    (x-nl)^2\theta[x-(n+1)l/2]\theta(x-nl)$(see Fig. $1$ (d)).
     Here $\theta(x)$ is the
    Heaviside step function and $l$ is the period of superlattice,
    $\gamma$ is a parameter characterizing the magnetic field strength.
    The above systems can be performed experimently$^{4,5,11}$.

    For a 2DEG subjected to a periodic magnetic field perpendicular
    to the $2$DEG plane,
the corresponding one-electron Hamiltonian reads
\begin{equation}
H={1\over 2m^*}[{\bf p}+e{\bf A(x)}]^2={1\over 2m^*}\{ p_x^2+[p_y+
e{\bf A(x)}]^2\},
\end{equation}
where $m^*$ is the effective mass of electron and $A(x)$ is the vector
potential in the Landau gauge.
Since $[P_y,H]=0$, the problem is translational invariant along the
$y$ direction.  Then the wave functions can be written
in the form ${1\over \sqrt{l_y}}e^{ik_yy}\psi(x)$,
where $k_y$ is the
wave-vector in the $y$ direction and $l_y$ the length of the magnetic
structure
in the $y$ direction.
By introducing the magnetic length $l_B=\sqrt{\hbar /e B_0}$ and the
cyclotron frequency $\omega_c=eB_0/m^*$, we  express
 the basic quantities in the dimensionless units:
 $(1)$ coordinates $\bf r \rightarrow
 {\bf r} l_B$. $(2)$ magnetic field $B(x)
 \rightarrow B(x)B_0$. $(3)$ the vector potential $A(x)\rightarrow A(x)B_0
 l_B$. $(4)$ the energy $E\rightarrow E\hbar\omega_c$. For $GaAs$ and an
 estimated $B_0=0.1T$ we have $l_B=813 nm$ , $\hbar\omega_c=0.17 meV$.
After some algebras, the following $1$D Schr{\"o}dinger
equation for $\psi(x)$ can be obtain$^4$
\begin{equation}
\left\{ {d^2\over dx^2}-[A(x)+k_y]^2+2E]
\right\}\psi(x)=0.
\end{equation}
The function $V(x,k_y)=[A(x)+k_y]^2$ can be interpreted as an
effective $k_y$-dependent electric potential. From this expression
we can
find out that, electron tunneling in MS is
inherently a complicated two dimensional process, which
depends on the electron's wave vectors in the
longitudinal and transverse directions of the $2$DEG, and thus possesses
no  general transmission splitting relation as in ES.

For the magnetic structure in region $[0, L=nl]$, we devide it into
$M(M>>1)$ segments, each of which has width $a=L/M$. The effective potential
in each segment can be viewed as constant and then the plane wave
approximation can be taken. In the $j$th segment, the wave functions
may be expressed as
\begin{equation}
\psi(x)=A_je^{ik_jx}+B_je^{-ik_jx}, x \in [ja,(j+1)a],
\end {equation}
where $k_j=\sqrt{{2E-[A(ja+a/2)+k_y]^2}}$, which may be either
real or pure imaginary.

  Without any loss of generality, we assume there is no magnetic field
in the incident and outgoing regions, then the wave functions can be
expressed by plane waves
\begin{eqnarray}
\left\{\begin{array}{ll}
               e^{ikx}+re^{-ikx}, & \mbox{$x<0$},
                 \\
            te^{ikx}, & \mbox{$x>L$},
            \end{array}\right.
\end{eqnarray}
where $k=\sqrt{2E-k_y^2}$ and $r$, $t$ are the reflection and transmission
amplitudes respectively.

The match of the wave functions and their derivatives at
$x=0$ and $x=L$ yields
\begin{equation}
\left[\begin{array}{c}1\\r\end{array}\right]
=\left[\begin{array}{cc} 1/2 & 1/(2ik)\\ 1/2 & -1/(2ik)\end{array}\right]
T_M\left[\begin{array}{cc} e^{ikL} & e^{-ikL}\\ ike^{ikL} & -ike^{-ikL}
\end{array}\right]
\left[\begin{array}{c}t\\0\end{array}\right],
\end{equation}
where
\begin{eqnarray}
T_M&=&\left[\begin{array}{cc} T_{11} & T_{12}\\
        T_{21} & T_{22}\end{array}\right]= \prod_{j=1}^{M}T^j_M \nonumber\\
   &=&\prod_{j=1}^{M}
\left[\begin{array}{cc} cos(k_ja) & -sin(k_ja)/k_j\\
k_jsin(k_ja) & k_jcos(k_ja)\end{array}\right],
\end{eqnarray}
where $T_M^j$ is the transfer matrix for the $j$th segment.

Transmission coefficient $T(E,k_y)$ for electron tunneling through
 the $n$-barrier
MS can be readily obtained from Eq. $(5)$
\begin{equation}
 T(E,k_y) =|t|^2 =\{1+(T_{11}^2+T_{22}^2+k^2T_{21}^2+T_{12}^2/k^2-2)/4\}^{-1}.
\end {equation}
With the transmission coefficient,
we calculate ballistic conductance from the well known
Landaur-B{\"u}ttiker formula$^{10}$
\begin{equation}
G/G_0=\int_{-\pi/2}^{\pi/2}T(E_F,\sqrt{2E_F}\sin{\theta})cos{\theta}
      d\theta ,
\end{equation}
where  $\theta$ is the
angle between the incidence velocity and the $x$ axis, $E_F$ is the
Fermi energy,
$G_0=e^2m^*v_Fl_y/\hbar^2$, and $v_F$ is the Fermi
velocity of electrons.

 First, ballistic conductances(in units of $G_0$) versus
  incidence energy in
 Kronig-Penny magnetic superlattice( KPMS ) were studied.
 Our results, shown in the left column of
 Fig. $2$, are calculated for the different number $n$ of magnetic
 barriers (which is also the number of the periods of magnetic superlattice
 except for $n=1$ case of half the magnetic period).
 The structure parameters are chosen to be $l=2, \gamma=2.5$ for solid
 curves and $l=2, \gamma=2$ for dashed curves. Let us inspect the
 conductance splitting at the magnetic strength $\gamma=2$.
 It is obvious that no
 resonant peak exists in the ballistic conductance
  for the single magnetic-barrier case.
One resonant peak is seen for double
 magnetic barriers and one sharper spike along with one resonant shoulder
 appears for triple magnetic barriers. With the increase of
 the number $n$ of magnetic
 barriers in KPMS, the total number of
 resonant conductance spikes and shoulders increases along with the
 resonant peaks and shoulders becoming sharper.
  As $n \rightarrow \infty$ , the peaks
 will fill in the energy windows of the magnetic minibands continuously
 as in the periodic ES$^7$.
 By counting the number of resonant peaks and resonant shoulders
 in  $n$-barrier KPMS, we found that, the number of resonant
 peaks and resonant shoulders, or the number of resonance splitting
 equals to $n-1$, which is
 the number of the magnetic barriers in KPMS.
 This is the corresponding $(n-1)$-fold resonant peak
 splitting for ballistic conductance
 in KPMS, which is similar to
 the $(n-1)$-fold resonant splitting for transmission
 in $n$-barrier ES.
 The splitting rules for ballistic conductance
 in KPMS is exactly the same as that for
 transmission in ES. With
 the magnetic strength $\gamma$ increasing,
 the resonant shoulders become resonant spikes and the
 resonant peaks are resolved more clearly. More importantly,
 the $(n-1)$-fold resonant peak splitting for ballistic conductance
 is unchanged.

      To find out the general rules for resonant peak splitting of
    ballistic conductance in MS
of arbitrary magnetic-barrier profile, we
calculated ballistic conductances versus incidence energy for
 Step MS in the middle column of Fig. $2$,
 Sinusoidal MS in the right column of Fig. $2$
  and  Sawtooth MS in Fig. $3$. The parameters
 for the calculated conductances of Step MS
 are the same as for the KPMS. While the parameters for
 Sinusoidal MS are set to be $l=2$, $\gamma=4$ for dashed curves and
  $l=2, \gamma=5$ for
 solid curves. In Fig. $3$, $l=2$, $\gamma=3$ are chosen
for dashed curves and $l=2$, $\gamma=3.5$ for solid curves.
 From Figs. $2$  and $3$, one can also observe clearly
 resonant splitting of the ballistic
 conductance peaks in Step MS, Sinusoidal MS and Sawtooth Ms.
 By checking the number
 of resonant peaks in ballistic conductances for
 Step MS, Sinusoidal MS and Sawtooth MS,
 we found that the number of resonant conductance peaks
 in the $n$-barrier Step MS,
 Sinusoidal MS  and Sawtooth MS is also $n-1$($n$ is the number of magnetic
 barriers).
 This indicates the
 existence of a general $n$-1 fold resonant splitting of conductance
 peaks in MS with $n$ identical magnetic barriers, which is independent
 of the magnetic-barrier profile.
 Then, We  can  generalize
 this rule as follows: for electron tunneling through
 magnetic superlattices with periodically arranged $n$ identical
 magnetic barriers,
 $(n-1)$-fold resonant peak splitting
 exists in ballistic conductance within each
 magnetic miniband. It is a general rule as the $n$-fold resonant peak
 splitting for transmission in $n$-electric-barrier superlattices.  For
 transmission of electron tunneling in magnetic
 superlattices,
 there is no such general splitting rule,
 since it is strongly dependent on the wave vector
 (momentum) normal to the tunneling direction.
 It is worth noting that the resonant peaks in ballistic
 conductances within lower energy minibands
 will be suppressed and that within higher energy minibands
 will be resolved gradually by the further-increased
 magnetic strength$^4$.

 As is well known, for electron tunneling through electric superlattice,
 when the incidence energy of electrons
 coincides with the energy of bound states
 in potential well, The resonant tunneling occurs (i.e.
 the transmission is $1$).
 Because of the coupling between the wells via tunneling through
 the barriers of finite width, the degenerate eigenlevels of the
 independent wells are split, consequently, these split levels redistribute
 themselves into groups aroud their unperturbed positions and form
 quasibands. This leads to the resonant splitting of transmission.
 As the
 number of periods(or the number of barriers) tends to infinity, the locally
 continuous energy distribution(energy band) is formed.
 Although electron tunneling in MS is more
 complicated than in ES due to its dependence on the
 perpendicular wave vector $k_y$(Ref. $4$), electron tunneling in MS
 is equivalent to that in ES for a given $k_y$ from
 the mathematical viewpoint. The resonant tunneling of
 electrons in MS results from
 the same physics as ES. We attributed the resonant peak splitting for
 ballistic conductance in magnetic superlattices to a collective
 effect of electron's wave-vector-dependent tunneling.
 Though the number of
 resonant transmission peaks in electron's tunneling through
 MS is closely related to the wave vector $k_y$ and may be
 different for different $k_y$, on an average, the
 number of resonant conductance peaks is the same as the number
 of wells in magnetic vector
 potential of MS. Since
 ballistic
 conductance is derived as the transmission averaged over
 all the possible wave vectors $k_y$, it
 can be viewed as the transmission of the electron's collective tunneling
 with a characteristic $k_y$
 through an average effective
 potential $V_{aver}(x)$, which has the same number of wells
 as the magnetic vector potential $A(x)$. This can be clearly seen if
 we plot the effective potential $V(x,k_y)$ as a function of
 $x$ and $k_y$ as did by Ibrahim et al.$^5$, the number
 of the main wells in the effective potential $V(x,k_y)$
 really equals to the number of the wells in the magnetic vector
 potential $A(x)$ on the whole and on average. Because the number
 of magnetic barriers in MS is $1$ more than the number of wells
 in the corresponding magnetic vector potential, as can be seen from
 Fig. $1$, $n$-1 fold resonant splitting occurs in the
 ballistic conductance peaks of $n$-barrier MS.

 In summary, we studied the resonant peak splitting effects
 for ballistic conductance
 in four kinds of magnetic superlattices of finite periods
 with identical
 magnetic barriers.  It is found that there is a general $(n-1)$-fold
 resonant peak splitting rule for
 ballistic conductance in $n$-identical-barrier magnetic
 superlattices, which is the analogy of $(n-1)$-fold transmission
 splitting in $n$-barrier electric superlattices.

  This work is supported by a key project for fundamental research
in the National Climbing Program of China.  One of us (Z. Y. Zeng)
acknowledges valuable discussions with Prof. L. M. Kuang and Prof.
G. J. Zeng during his visit to Hunan Normal University.

\vspace{.5cm}
\noindent
{\bf References}

\vspace{.1cm}
\noindent
\hspace{0.2cm} $\ddagger$ E-mail address: zyzeng@mail.issp.ac.cn
\begin{enumerate}
\item P. Vasilopoulos and F. M. Peeters, Superlatt. Microstruct.
{\bf 4}, 393 (1990);
D. P. Xue and G. Xiao, Phys. Rev. B. {\bf 45},
5986 (1992);
 F. M. Peeters and P. Vasilopoulos, Phys. Rev. B.
{\bf 47}, 1446 (1993).
\item H. A. Carmona et al., Phys. Rev. Lett. {\bf 74}, 3009 (1995);
P. D. Ye et al., ibid. {\bf 74}, 3013 (1995).
\item F. M. Peeters and A. Matulis, Phys. Rev. B. {\bf 48}, 15166 (1993)
\item A. Matulis, F.M. Peeters, and P. Vasilopoulos, Phys. Rev. Lett.
{\bf 72}, 1518 (1994);
 S. S. Allen and S. L. Richardson, Phys. Rev. B {\bf 50}, 11693 (1994);
 J. Q. You, Lide Zhang, and P.K.Ghosh, Phys. Rev. B {\bf 52}, 17243 (1995);
Yong Guo, Binlin Gu, and Wenhui Duan, Phys. Rev. B {\bf 55}, 9314 (1997).
\item  L. S. Ibrahim and F. M. Peeters, Phys. Rev. B. {\bf 52}, 17321 (1995);
A. Krakovsky, Phys. Rev. B. {\bf 53}, 8469 (1996).
\item R. Tsu and L. Esaki, Appl. Phys. Lett. {\bf 22}, 562 (1973)
\item Xue-Wen Liu and A. P. Stamp, Phys. Rev. B. {\bf 50}, 588 (1994)
\item Yong Guo, Bin-lin Gu, Zhi-Qiang Li, Jing-Zhi Yu, and Yoshiyuk. Kawazoe,
 J. Appl. Phys. {\bf 83}, 4545 (1998).
\item J. Q. You and Lide Zhang, Phys. Rev. B {\bf 54}, 1526 (1996)
\item B{\"u}ttiker, Phys. Rev. Lett. {\bf 57}, 1761 (1986)
\item
 P. D. Ye, D. Weiss, R. R. Gerhardts, M. Seeger, K. Von. Klizing,
 K.Eberl, and H. Nickel, Phys. Rev. Lett. {\bf 74}, 3013 (1995).
\end{enumerate}

\newpage
\noindent
{\bf Figure Captions}

{\bf Fig.~1}~ The magnetic field profiles and the corresponding
vector potential about four kinds of magnetic superlattices, where
only $3$ periods are plotted.

\vspace{.5cm}
\noindent
{\bf Fig.~2}~ Ballistic conductances for KPMS
, Step MS and Sinusoidal MS. The left column of
the figure corresponds to the KPMS, the middle
column to Step MS and the right column to Sinusoidal MS.
Here $l=2$, $\gamma=2$ for dashed curves and $\gamma
=2.5$ for solid curves in KPMS and Step MS cases, while
 $\gamma=4$ for dashed curves and   $\gamma=5$ for
solid curves in Sinusoidal MS case. $n$ is the number of
the magnetic barriers(also the number of magnetic period except
for n=1 case of half the magnetic period).
.
\vspace{.5cm}
\noindent

{\bf Fig.~3}~ Ballistic conductance for
 Sawtooth MS.  Here $l=2$,  $\gamma=3$ for dashed curves and
 $\gamma=3.5$ for
solid curves.
\end{document}